\documentclass[twoside]{ilcws08}
\usepackage[latin1]{inputenc}
\usepackage[dvips]{graphicx,epsfig,color}
\usepackage{wrapfig,rotating}
\usepackage{amssymb,amsmath,array}

\begin{document}
\title{
ILC Instrumentation R\&D at SCIPP} 
\author{J. Carman$^1$, S. Crosby$^1$, V. Fadeyev$^1$, R. Partridge$^2$, B. A. Schumm$^1$,
N. Spencer$^1$, and M. Wilder$^1$
\vspace{.3cm}\\
1- Santa Cruz Institute for Particle Physics \\
Santa Cruz, CA 95062,  U.S.A.
\vspace{.1cm}\\
2- SLAC National Accelerator Laboratory \\
Stanford, CA, 94025, U.S.A. \\
}

\maketitle

\begin{abstract}
The Santa Cruz Institute for Particle Physics (SCIPP) continues to be
engaged in research and development towards an ILC detector. The latest
efforts at SCIPP are described, including those associated with the LSTFE
front-end readout ASIC, the use of charge division to obtain a longitudinal
coordinate from silicon strip detectors, and the contribution of strip
resistance to readout noise.
\end{abstract}

\section{The LSTFE Time-Over-Threshold Readout ASIC}

\begin{figure}
\centerline{\includegraphics[width=0.50\columnwidth]{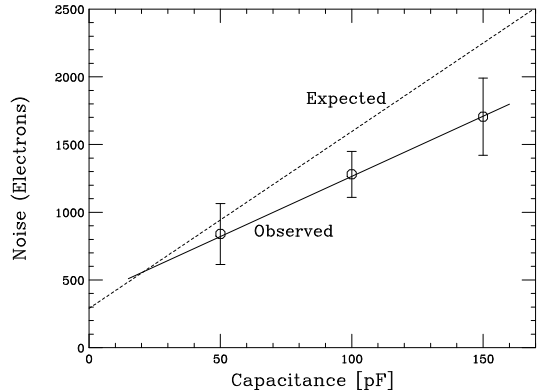}}
\caption{Measured vs. expected noise for the LSTFE-I prototype ASIC,
in equivalent electrons vs. capacitive load in pF.}\label{Fig:noise}
\end{figure}

The LSTFE ASIC features
a long ($\sim$ 1.5 $\mu$s) shaping time, typical for ILC 
silicon sensor readout applications,
that limits readout noise from capacitive and series-resistance
load. After amplification and shaping, the signals are split
in two and directed to two separate comparators. One runs
at a high ($\sim$ 1.2 fC) threshold to suppress noise, while the
other runs at a low ($\sim$ 0.4 fC) threshold to maximize the
information used in constructing the centroid of pulses
that trigger the high threshold. The charge-amplitude
measurement is provided by the duration of low-threshold 
comparator's time-over-threshold, which, in a full 
implementation of the ASIC, would be stored digitally in
an on-board FIFO and read out asynchronously. The chip
includes a power-cycling feature that is designed to
allow the chip to be powered on in 1 msec, allowing 
the chip to reduce its power consumption by 99\%
by exploiting the 5 Hz duty cycle of the Linear
Collider. Figure~\ref{Fig:noise} shows the observed 
noise performance of the LSTFE-I prototype ASIC, as
a function of capacitive load. The best-fit line yields
  $$\sigma_{e} = 375 + 8.9 \cdot C,$$
in equivalent electrons
per pF.

A refined LSTFE-II prototype has been fabricated,
and is being readied for testing at the SCIPP laboratory.
Due to unanticipated leakage currents, the switch-on
time of the LSTFE-I was degraded to 30 msec; this has
been addressed in the design of the LSTFE-II by adding low-power
feedback loops that cancel leakage currents during
the nominal power-off phase. The design includes an
additional intermediate amplification stage that
should allow further improvements in signal-to-noise
and channel-to-channel matching. The return-to-baseline
has been hastened through a re-configuration of the
pulse-shaping feedback circuitry, which should improve
the time-over-threshold resolution. The signal-to-noise
performance has been optimized for an 80 cm ladder
(approximately 100 pF load), which is more likely
for an ILC long-ladder application than the original
target of 167 cm. The chip has 128 channels (256 
comparators) whose outputs are sampled at 3 MHz, and
multiplexed at 32:1 into eight 100 MHz LVDS outputs.
The group is planning to use the LSTFE-II in a test-beam
run in mid-2009.

\section{Charge Division with Resistive Sensors}

We have begun to explore the possibility of using charge
division across a resistive readout channel~\cite{owen,radeka,pullia} to obtain
information about the longitudinal coordinate of the
deposition of minimum-ionizing particles. To this end,
the SiD sensor fabrication included a short (5 cm) test
sensor for which there was supposed to be direct
connectivity with the resistive implant (total 
resistance specified to be 600 k$\Omega$), with no metal strip
connecting the readout pads at the near and far ends of the
implant. Due to an error in the fabrication process,
the metal strip was included, with essentially continuous
connectivity to the length of the implant, rendering resistive
charge division studies impossible. 

Instead, a hardware simulation of the resistive
readout was developed, using a PC board loaded with
discrete components that approximate the distributed
RC network of the planned SiD charge division sensor
prototype. The total resistance of 600 $k\Omega$,
and an assumed capacitance of 1.2 pF/cm, were
divided into 10 discrete sections, with either
end read out by commercial TI OPA657 low-noise FET 
operational amplifiers. Two later amplification stages,
making use of Analog Devices ADA4851 rail-to-rail video amps,
were used to shape the readout pulse and
provide a rise-time of 2.0 $\mu$s. 

Injecting charge at the circuit's discrete nodes
produced the expected linear charge division
between the near- and far-end amplifiers, to
within measurement errors. By injecting charge through
the introduction of a voltage step across 
a known capacitance, the gain was calibrated, allowing
an expression of readout noise in terms of equivalent charge.
Table~\ref{tab:div_noise} shows the equivalent-charge
readout noise for several different readout
configurations.

\begin{table}
\centerline{\begin{tabular}{|l|c|}
\hline
Configuration  & Electron-Equivalent Readout Noise \\\hline 
Nominal  & 0.64 fC  \\
Ground near-end input &  0.26 fC   \\
Disconnect far-end amp; ground network at that end & 0.66 fC  \\
Disconnect far-end amp; float network at that end &  0.44 fC  \\
\hline
\end{tabular}}
\caption{Equivalent-electron readout noise from the near-end
amplifier, for various configuration of the discrete network.}
\label{tab:div_noise}
\end{table}

The value of 0.26 fC obtained with the input to the near-end
amplifier grounded suggests that the level of noise introduced
by the commercial preamplifier, while not ideal, is not
dominant when the resistive strip is read out. Since the
dynamic impedance of the amplifier chain is expected to be
small relative to the network load resistance, one would 
expect little difference in noise if the far end amplifier input is
grounded, and the amplifier on that end is disconnected; that
was indeed observed. However, the reduction in noise
that was observed when the far end of the network was
disconnected and floated is not understood. 

To explore
this further, the correlation between the near- and far-end amplifier
noise was measured, which is expected to show anti-correlation
if dominated by the resistive component of the load network~\cite{radeka}.
A correlation coefficient consistent with zero was measured;
if instead, the network was replaced by a single 600 $k\Omega$
resistor, a correlation coefficient of $\rho = -0.51$
was measured. This suggests that RC network effects may
be reducing the correlation between the near- and far-end
readout noise, which would in fact be beneficial for
the precision of the charge-division measurement (see below). A
SPICE simulation of the network and readout is being developed,
so that the behavior can be understood more rigorously.

If $N,F$ are the near- and far-end signals, respectively,
than the fractional longitudinal coordinate measurement
is given by $f = (N-F)/(N+F)$, with an uncertainty
$$ df = -{2 \over (1+x)^2}dx  $$
$$ x = N/F     $$
$$ dx = x \Bigl[ {dN \over N} - {dF \over F} \Bigr] , $$
from which it can be seen that anti-correlation
between the near- and far-end amplifiers would
worsen the longitudinal coordinate measurement.
Using the measured noise characteristics, assuming
no correlation between the two ends, and a 4 fC deposition
in the center of the network, a fractional resolution of
0.23 (2.3 cm for a 10 cm sensor) is estimated. While this
is only moderately better than $1/\sqrt{12} \simeq 0.29$,
there is still much to be done in terms of understanding
and optimizing the charge-division technique.

\section{Microstrip Sensor Readout Noise for ILC Applications}

Readout noise depends upon many factors, including
series and parallel resistive loads ($R_S$, $R_B$), 
series and parallel amplifier noise ($e_{na}$, $i_{na}$), 
capacitive load (C), and leakage
current ($I_d$). According to~\cite{spieler}, these factors
contribute to the overall equivalent-electron noise as
$$ Q^2 = F_i \tau (2eI_d + {4kT \over R_B} + i_{na}^2) +
  {F_v C^2 \over \tau}(4kT R_S + e_{na}^2) + 4 F_v A_f C^2, $$
where $\tau$ is the rise time, $F_i$ and $F_v$ are signal shape
parameters of order one that characterize the amplifier
response to current and voltage noise sources respectively, 
and $A_f$ is the contribution from
$1/f$ noise. 

With its emphasis on precision, ILC silicon strip applications
tend to employ narrow and/or long strips, which leads to 
significant series resistance. 
Taken at face value, this expression suggests that the
readout noise is likely to be limited by the series noise
contribution, even at the relatively large shaping times
expected for ILC applications. 

The SCIPP ILC group has been interested in exploring the
series contribution to readout noise, and developing
mitigating approaches (such as reading the ladder out
from its center rather than from either end) to reduce
the series noise contribution should it be dominant.
Initial studies done with a long ladder, read out
by the LSTFE-I prototype, and composed of
auxiliary sensors from the GLAST production run that
have a strip width of 65 $\mu$m, showed roughly the
expected dependence of readout noise on ladder length.
However, even at the longest ladder length that was
tested (143 cm), the expectation from~\cite{spieler}
is that series resistive noise was only about 20\% greater than
the contribution from the purely capacitive term, so it was
difficult to be confident about the size of the contribution
from series noise. In addition, reading the ladder out
from the center, which might be expected to reduce the
series noise contribution by a factor of two, produced
no appreciable reduction in noise.

To explore this further, the group has attempted to build
a long ladder out of unused CDF Layer00 sensors, which have
a strip resistance of 40 $\Omega$/cm, approximately three
times that of the GLAST sensors, and which should thus
exhibit readout noise dominated by series resistance
for ladder lengths of 50 cm or more. Because these
sensors were intended for short shaping-time applications,
the bias resistance used (approximately 0.6 M$\Omega$) would
dominate if read out by the LSTFE-I. Thus, the bias
resistors were severed with a focussed laser beam, and 
several strips were biased with 30 M$\Omega$ through
a micro-jig developed at SCIPP. For a single CDF Layer00
sensor, the group observes $\sim$20\% greater noise than
expected from experience with the GLAST sensors, and is
working to identify and eliminate the extraneous contribution.
The group hopes to soon have results from a longer, 
series-resistance-dominated ladder.

\section{Summary}

SCIPP continues to make headway in R\&D areas specific to the 
application of silicon strip detectors to the ILC (LSTFE
ASIC development) as well as in areas with more generic
uses (charge division, series resistance contribution to readout
noise). The group hopes to have new results soon on the
refined LSTFE-II prototype chip, which will be suitable for
use in a test-beam run, as well an improvement of the understanding of
series noise. The group will also continue to optimize the
charge-division strategy, with hopes of finding an operating
strategy that significantly improves on the intrinsic $1/\sqrt{12}$
longitudinal resolution while preserving the performance in the
dimension transverse to the strips.




\begin{footnotesize}



%

\end{footnotesize}


\end{document}